# Electromagnetic Cascade in the Early Universe and its Application to the Big-Bang Nucleosynthesis

M. Kawasaki[1] and T. Moroi[2][†]

[1] *Institute for Cosmic Ray Research, The University of Tokyo, Tanashi 188, Japan*

[2] *Department of Physics, Tohoku University, Sendai 980-77, Japan*

## Abstract

We investigate the electromagnetic cascade initiated by injection of very high energy photons in the early Universe and calculate the cascade spectrum by solving a set of Boltzmann equations numerically. In the calculation we take account of Compton scattering off the background electrons and pair creation off the background nucleons as well as photon-photon processes and inverse Compton scattering. We also apply our cascade spectrum to the big bang nucleosynthesis with photo-dissociation processes due to heavy unstable particles and obtain the constraint on their lifetime and abundance.

*Subject headings* : cosmology–gamma rays:general– elementary particles

Email: kawasaki@ctsun1.icrr.u-tokyo.ac.jp, moroi@tuhep.phys.tohoku.ac.jp

---

[†]Fellow of the Japan Society for the Promotion of Science.

# I. Introduction

Electromagnetic cascade initiated by very high energy electrons and photons is very important in a number of astrophysical situations (e.g., Zdziarski 1988 and references therein). For example, very high energy $\gamma$-rays emitted from Cyg X-3 propagate through the cosmic microwave background and cause the electromagnetic cascade. Electromagnetic cascade also expected in active galactic nuclei, compact sources and galactic halos. In the early Universe the high energy photons (and electron-positrons) might be produced by some exotic sources such as primordial black holes and unstable heavy particles.

The high energy photons emitted from the above sources scatter off the background photon and produce electron-positron pairs if their energy $\epsilon_\gamma$ is greater than $m_e/\bar{\epsilon}_\gamma$ ($m_e$: electron mass, $\bar{\epsilon}_\gamma$: energy of the background photon). Then the electrons and positrons produce the high energy photons by inverse Compton scatterings. This cascade process continues until the energy of the photons becomes too low to produce electron-positron pairs. When relevant processes are electron-positron pair production by photon-photon interaction and inverse Compton scattering, the spectrum of the cascade photons are found to be $\propto \epsilon_\gamma^{-1.5}$ (e.g., Berezinskiĭ 1990). The spectrum is modified by the scattering between the high energy photons and the background photons as noticed by Svensson and Zdziarski (1990), and the spectrum obtained by Svensson and Zdziarski (1990) is good enough for most of astrophysical applications. However, when one study the electromagnetic cascade in the early Universe ($z \gtrsim 10^3$), one should take other processes into account. In the early Universe, the number densities of electrons and nucleons are so high that the Universe is opaque to photons by Compton scattering off the background electrons and pair creation by the background nucleons. These Compton scattering and pair creation are main energy-loss processes for the photons whose energy is too low to produce electron-positron pairs by scattering off the background photons. Therefore these processes which are neglected in the previous work play an important role to determine the low-energy region of the cascade spectrum.

In cosmological applications, it is essential to know the precise form of the cascade spectrum. One of the most important cosmological effects due to electromagnetic cascade is the photo-dissociation of light elements (D, $^3$He, $^4$He). It is well established that D and He are synthesised in the early Universe when the temperature is between 1MeV and 10keV. The theoretically predicted abundances of these light elements are in good agreement with the observed ones. However this success of the big bang nucleosynthesis (BBN) is destroyed if the high energy photons produced in the cascade destroy the light elements. By considering this effect, constraints are imposed on energy sources such as massive neutrinos (e.g. Kolb and Turner 1990 and references therein). When one places these constraints, it is necessary to use the precise cascade spectrum.

In this paper we study the electromagnetic cascade in the early Universe and obtain the spectrum of the high energy photons and electrons including the interaction with background electrons and nucleons (*i.e.* Compton scattering and pair creation). We have already applied our spectrum to the specific problems (Kawasaki and Moroi 1994a, 1994b). Here we describe how the spectrum is obtained in detail and apply the spectrum to BBN with general unstable heavy particles.

Recently, Protheroe, Stanev and Berezinsky (1994) studied the similar problem by Monte Carlo calculation. Our method to obtain the spectrum is different from theirs and may be much more simple since our method is just to find steady-state-solution to the Boltzmann equations numerically. We also present some simple formula which fit to our cascade spectrum.

In Section II the relevant processes in electromagnetic cascade are reviewed. In Section III the method to obtain the cascade spectrum is described in detail. The application to BBN and constraint on the general decaying particle are given in Section IV. Section V is devoted to conclusions.

## II. Electromagnetic Cascade Processes

Once high energy photons are emitted into the thermal bath in the early universe, they induce electromagnetic cascade processes. At first, we briefly review the elementary processes which determine the shape of the spectrum.

- If the energy of the high energy photon is sufficiently large, the photon-photon pair creation process ($\gamma + \gamma_{\rm BG} \to e^+ + e^-$, where $\gamma_{\rm BG}$ represents the background photon) is the most dominant process since the cross section or the number density of the target is much larger than other processes. However, this process has threshold, *i.e.* the energy of the initial high energy photon has to be larger than $m_e^2/\bar{\epsilon}_\gamma$ with $\bar{\epsilon}_\gamma$ being the energy of the background photon. As we will see later, this process determines the shape of the spectrum of the high energy photon for $\epsilon_\gamma \gtrsim m_e^2/22T$.

- Below the effective threshold of the double photon pair creation, high energy photons lose their energy by the photon-photon scattering ($\gamma + \gamma_{\rm BG} \to \gamma + \gamma$). But in the limit of $\epsilon_\gamma \to 0$, the total cross section for the photon-photon scattering is proportional to $\epsilon_\gamma^3$ and this process loses its significance.

- Finally, photons lose their energy by scattering off charged particles in the thermal bath. The dominant processes are pair creation in the nuclei ($\gamma + N \to e^+ + e^- + N$, where $N$ represents the nuclei in the background), and Compton scattering off the thermal electron ($\gamma + e_{\rm BG}^\pm \to \gamma + e^\pm$, where $e_{\rm BG}^\pm$ represents the electron and positron in the background).

- Emitted high energy electrons and positrons lose their energy by the inverse Compton scattering off the background photon ($e^{\pm} + \gamma_{\rm BG} \to e^{\pm} + \gamma$).

Taking these processes into account, the Boltzmann equations for the photon and the electron distribution functions $f_\gamma$ and $f_e$ can be written as

$$\frac{\partial f_\gamma(\epsilon_\gamma)}{\partial t} = \left.\frac{\partial f_\gamma(\epsilon_\gamma)}{\partial t}\right|_{\rm DP} + \left.\frac{\partial f_\gamma(\epsilon_\gamma)}{\partial t}\right|_{\rm PP} + \left.\frac{\partial f_\gamma(\epsilon_\gamma)}{\partial t}\right|_{\rm PC} + \left.\frac{\partial f_\gamma(\epsilon_\gamma)}{\partial t}\right|_{\rm CS}$$
$$+ \left.\frac{\partial f_\gamma(\epsilon_\gamma)}{\partial t}\right|_{\rm IC} + \left.\frac{\partial f_\gamma(\epsilon_\gamma)}{\partial t}\right|_{\rm S} + \left.\frac{\partial f_\gamma(\epsilon_\gamma)}{\partial t}\right|_{\rm EXP}, \qquad (1)$$

$$\frac{\partial f_e(\epsilon_e)}{\partial t} = \left.\frac{\partial f_e(\epsilon_e)}{\partial t}\right|_{\rm DP} + \left.\frac{\partial f_e(\epsilon_e)}{\partial t}\right|_{\rm PC} + \left.\frac{\partial f_e(\epsilon_e)}{\partial t}\right|_{\rm CS} + \left.\frac{\partial f_e(\epsilon_e)}{\partial t}\right|_{\rm IC}$$
$$+ \left.\frac{\partial f_e(\epsilon_e)}{\partial t}\right|_{\rm S} + \left.\frac{\partial f_e(\epsilon_e)}{\partial t}\right|_{\rm EXP}, \qquad (2)$$

where DP (PP, PC, CS, IC, S, and EXP) represents double photon pair creation (photon-photon scattering, pair creation in nuclei, Compton scattering, inverse Compton scattering, the contribution from the high energy photon source, and the effects of the expansion of the Universe). Full details are given in Appendix.

Our purpose is to solve the above Boltzmann equations (1) and (2), and to obtain the high energy photon and electron spectra. In eqs. (1) and (2), there are two time scales which govern the evolution of the spectra, *i.e.* the one is that of the electromagnetic processes and the other is the Hubble time. Since the former is much shorter than the latter, it is difficult to determine the evolution of the spectra by the numerical integration of the full Boltzmann equations. In order to solve eqs.(1) and (2), we use another method. When the photon source is active, the time variation of the source is small in one Hubble time $H^{-1}$ for most cases. Then, the photon source can be regarded as stationary and the stationary state is established by the balance between absorption and emission processes at each epoch. Based on this fact, we have obtained the photon and electron spectra from the stationary solutions to the Boltzmann equations. In fact, to derive a stationary solution is much more practical than the full integration of the Boltzmann equations (1) and (2). In the next section, we will explain the basic methods which we have used.

## III. Numerical Integration of the Boltzmann Equation

In deriving high energy photon and electron spectra, we have to use numerical method since the Boltzmann equations (1) and (2) are very much complicated. In this section, we explain our approach to the Boltzmann equations.

At first, we classify the right-hand side of eqs.(1) and (2) into the "outgoing" parts and "incoming" ones. Below, we ignore the expansion terms $(\partial f_\gamma/\partial t)|_{\rm EXP}$ and $(\partial f_e/\partial t)|_{\rm EXP}$ in the Boltzmann equations (1) and (2) since the expansion rate of the Universe is much smaller than the scattering rates of electromagnetic processes. Then, eqs.(1) and (2) can be written as

$$\frac{\partial f_\gamma(\epsilon_\gamma)}{\partial t} = -\Gamma_\gamma(\epsilon_\gamma;T)f_\gamma(\epsilon_\gamma) + \dot{f}_{\gamma,\text{IN}}(\epsilon_\gamma), \tag{3}$$

$$\frac{\partial f_e(\epsilon_e)}{\partial t} = -\Gamma_e(\epsilon_e;T)f_e(\epsilon_e) + \dot{f}_{e,\text{IN}}(\epsilon_e), \tag{4}$$

where "incoming" terms $\dot{f}_{\gamma,\text{IN}}$ and $\dot{f}_{e,\text{IN}}$ can be obtained as the sums of the contributions from the source terms and the functionals of the distribution functions of the photon and the electron;

$$\begin{aligned}
\dot{f}_{\gamma,\text{IN}}(\epsilon_\gamma) &= \left.\frac{\partial f_\gamma(\epsilon_\gamma)}{\partial t}\right|_S + \int_{\epsilon_\gamma}^\infty d\epsilon'_\gamma K_{\gamma,\gamma}(\epsilon_\gamma,\epsilon'_\gamma;T)f_\gamma(\epsilon'_\gamma) \\
&+ \int_{\epsilon_\gamma}^\infty d\epsilon'_e K_{\gamma,e}(\epsilon_\gamma,\epsilon'_e;T)f_e(\epsilon'_e),
\end{aligned} \tag{5}$$

$$\begin{aligned}
\dot{f}_{e,\text{IN}}(\epsilon_e) &= \left.\frac{\partial f_e(\epsilon_e)}{\partial t}\right|_S + \int_{\epsilon_e}^\infty d\epsilon'_\gamma K_{e,\gamma}(\epsilon_e,\epsilon'_\gamma;T)f_\gamma(\epsilon'_\gamma) \\
&+ \int_{\epsilon_e}^\infty d\epsilon'_e K_{e,e}(\epsilon_e,\epsilon'_e;T)f_e(\epsilon'_e).
\end{aligned} \tag{6}$$

The explicit forms of $\Gamma_A$ and $K_{AB}$ ($A,B = \gamma,e$) can be obtained from the full details of the Boltzmann equations given in Appendix.

As mentioned in the previous section, the strategy we use here is to calculate stationary solutions to the Boltzmann equations (1) and (2), which obey

$$\frac{\partial f_\gamma(\epsilon_\gamma)}{\partial t} = \frac{\partial f_e(\epsilon_e)}{\partial t} = 0. \tag{7}$$

In this case, the distribution functions $f_\gamma$ and $f_e$ of the photon and the electron can be formally given as

$$f_\gamma(\epsilon_\gamma) = \frac{\dot{f}_{\gamma,\text{IN}}(\epsilon_\gamma)}{\Gamma_\gamma(\epsilon_\gamma;T)}, \tag{8}$$

$$f_e(\epsilon_e) = \frac{\dot{f}_{e,\text{IN}}(\epsilon_e)}{\Gamma_e(\epsilon_e;T)}. \tag{9}$$

The important thing is that the "incoming" terms $\dot{f}_{\gamma,\text{IN}}(\epsilon_\gamma)$ and $\dot{f}_{e,\text{IN}}(\epsilon_e)$ depend only on $f_\gamma(\epsilon')$ and $f_e(\epsilon')$ with $\epsilon' > \epsilon_\gamma, \epsilon_e$. Therefore, if the distribution functions $f_\gamma(\epsilon_\gamma)$ and $f_e(\epsilon_e)$ with $\epsilon_\gamma, \epsilon_e > \epsilon$ (and the source terms) are known, we can obtain $f_\gamma(\epsilon)$ and $f_e(\epsilon)$ from eqs.(8) and (9). By using this fact, we have solved the Boltzmann equations (1) and (2) with monochromatic high energy photon injection. As one can see, extensions to the cases with non-monochromatic sources or the high energy electron source are trivial.

In our numerical calculations, we have used the mesh points $E_i$ ($0 \leq i \leq N_{\mathrm{mesh}} + 1$) with $E_0 = \epsilon_{\mathrm{min}}$ and $E_{N_{\mathrm{mesh}}} = \epsilon_0$, where $\epsilon_{\mathrm{min}}$ is some minimum energy we are concerning (which we take $\epsilon_{\mathrm{min}} = 1\mathrm{MeV}$) and $\epsilon_0$ is the energy of the primary injecting photons. The energy range between $\epsilon_{\mathrm{min}}$ and $\epsilon_0$ is divided logarithmically, i.e. the $i$-th mesh point $E_i$ is given by

$$E_i = \epsilon_{\mathrm{min}} \left( \frac{\epsilon_0}{\epsilon_{\mathrm{min}}} \right)^{i/N_{\mathrm{mesh}}}. \tag{10}$$

By using the fact that the incoming photons into the mesh point $E_{N_{\mathrm{mesh}}}$ are supplied only by the photon source, incoming terms for $i = N_{\mathrm{mesh}}$ is given by

$$\dot{f}_{\gamma,\mathrm{IN}}(E_{N_{\mathrm{mesh}}}) \simeq \dot{n}_\gamma \times \frac{1}{\Delta_{N_{\mathrm{mesh}}}}, \tag{11}$$

$$\dot{f}_{e,\mathrm{IN}}(E_{N_{\mathrm{mesh}}}) \simeq 0, \tag{12}$$

where $\dot{n}_\gamma$ is the production rate of monochromatic photons, and $\Delta_i \equiv (E_{i+1} - E_{i-1})/2$. Combining eqs.(11) and (12) with eqs.(8) and (9), we can obtain the distribution functions at the $N_{\mathrm{mesh}}$-th mesh point $f_\gamma(E_{N_{\mathrm{mesh}}})$ and $f_e(E_{N_{\mathrm{mesh}}})$.

Next we determine the distribution functions for lower energy region. Essentially, "incoming" terms for the $i$-th mesh point are derived from the distribution functions $f_\gamma(E_j)$ and $f_e(E_j)$ with $j > i$. Discretizing eqs.(5) and (6), "incoming" terms are (approximately) given by

$$\dot{f}_{\gamma,\mathrm{IN}}(E_i) \simeq \sum_{j>i} \Delta_j \left\{ K_{\gamma\gamma}(E_i, E_j) f_\gamma(E_j) + K_{\gamma e}(E_i, E_j) f_e(E_j) \right\}, \tag{13}$$

$$\dot{f}_{e,\mathrm{IN}}(E_i) \simeq \sum_{j>i} \Delta_j \left\{ K_{e\gamma}(E_i, E_j) f_\gamma(E_j) + K_{ee}(E_i, E_j) f_e(E_j) \right\}, \tag{14}$$

from which we can obtain $f_\gamma(E_i)$ and $f_e(E_i)$ by using eqs.(8) and (9).[1] In a way explained above, we have calculated the photon and electron distribution functions $f_\gamma(E_i)$ and $f_e(E_i)$ at each mash points from $i = N_{\mathrm{mesh}} - 1$ to $i = 0$ in order.

In Figs.1–3, we have shown the photon spectra for the cases $\epsilon_{\gamma 0} = 10\mathrm{TeV}$, $1\mathrm{TeV}$, $100\mathrm{GeV}$ and $T = 1\mathrm{eV}$, $100\mathrm{eV}$. As we will see, distribution function $f_\gamma(\epsilon_\gamma)$ with $\epsilon_\gamma \gtrsim m_e^2/22T$ is very much small compared with that of the low energy region, since the double photon pair creation process is effective for such high energy region. As we mentioned before, the scattering rate of the double photon pair creation process is much larger than those for other cascade processes (if kinematically allowed), and hence the photon spectrum are extremely suppressed for the region $\epsilon_\gamma \gtrsim m_e^2/22T$. Below this effective threshold, the double photon pair creation process receives a Boltzmann suppression and other processes become important. In fact, the photon-photon scattering is the most dominant process just below the threshold, while the shape of the photon spectrum for the sufficiently low energy region is determined by the scattering processes off the charged particles in the background.

---

[1] In fact, eqs.(11) – (14) receive corrections of the order of $\Delta$, which is due to the discretization of the energy range. In our numerical calculations, we have taken these corrections into account.

As one can see in Figs.1–3, both the photon spectrum for sufficiently low energy region and that for the photon-photon scattering region obey power-law spectrum; $f_\gamma(\epsilon_\gamma) \propto \epsilon_\gamma^P$. We fit the photon spectrum at the sufficiently low energy region ($\epsilon_\gamma \lesssim m_e^2/80T$) as

$$f_\gamma(\epsilon_\gamma) \simeq N_{\rm low} \left(\frac{\dot{\rho}_{\rm IN}}{\rm GeV^5}\right) \left(\frac{T}{\rm GeV}\right)^{-3} \left(\frac{\epsilon_\gamma}{\rm GeV}\right)^{P_{\rm low}}, \qquad (15)$$

and that for the photon-photon scattering region ($m_e^2/80T \lesssim \epsilon_\gamma \lesssim m_e^2/22T$) as

$$f_\gamma(\epsilon_\gamma) \simeq N_{\rm pp} \left(\frac{\dot{\rho}_{\rm IN}}{\rm GeV^5}\right) \left(\frac{T}{\rm GeV}\right)^{-6} \left(\frac{\epsilon_\gamma}{\rm GeV}\right)^{P_{\rm pp}}, \qquad (16)$$

with $\dot{\rho}_{\rm IN}$ being the total amount of the energy injection from the high energy photon source;

$$\dot{\rho}_{\rm IN} = \int d\epsilon_\gamma \epsilon_\gamma \left.\frac{\partial f_\gamma(\epsilon_\gamma)}{\partial t}\right|_{\rm S}. \qquad (17)$$

The amplitude of the spectrum is proportional to the mean free time of the photon. In low energy region the mean free time is determined by Compton scattering and depends on the background temperature as $\sim T^{-3}$. For the photon-photon scattering region, the amplitude is proportional to $\sim T^{-6}$ since the cross section depends on $\sim T^3$. For the cases $T = 1{\rm eV}, 10{\rm eV}, 100{\rm eV}$ and $\epsilon_0 = 10{\rm TeV}, 1{\rm TeV}, 100{\rm GeV}, 10{\rm GeV}$, we have calculated the fitting parameters $P_{\rm low}$, $P_{\rm pp}$, $N_{\rm low}$ and $N_{\rm pp}$, and the results are shown in Table 1.[2] These variables slightly depend on the background temperature, while their dependence on the initial photon energy $\epsilon_0$ is insignificant.

Before closing section, we comment on the case with high energy electron sources. By numerical calculations, we have checked that the spectra in the case with high energy electron injection are almost the same as those with photon injection if the background temperature and the total amount of the energy injection are fixed. Numerically, their differences are at most $O(10\%)$. Therefore, the formulae given in eqs.(15) and (16) with Table 1 are well approximated ones if one redefines $\dot{\rho}_{\rm IN}$ given in eq.(17) as

$$\dot{\rho}_{\rm IN} = \int d\epsilon_\gamma \epsilon_\gamma \left.\frac{\partial f_\gamma(\epsilon_\gamma)}{\partial t}\right|_{\rm S} + \int d\epsilon_e \epsilon_e \left.\frac{\partial f_e(\epsilon_e)}{\partial t}\right|_{\rm S}. \qquad (18)$$

# IV BBN and Photo-Dissociation of Light Elements

The injection of high energy photons might destroy the success of BBN since the photons produced through the electromagnetic cascade destroy the light elements (D, $^3$He, $^4$He).

---

[2] Notice that the formulae (15) and (16) are relevant for the cases when the initial photon energy $\epsilon_0$ is much larger than the effective threshold of the double photon pair creation process.

When the energy of the high energy photon is relatively low, *i.e.* $2\text{MeV} \lesssim \epsilon_\gamma \lesssim 20\text{MeV}$, the D, T and $^3$He are destroyed and their abundances decrease. On the other hand, if the photons have high energy enough to destroy $^4$He, it seems that such high energy photons only decrease the abundance of all light elements. However since D, T and $^3$He are produced by the photo-dissociation of $^4$He whose abundance is much higher than the other elements, their abundances can increase or decrease depending on the number density of the high energy photon. When the number density of the high energy photons with energy greater than $\sim 20\text{MeV}$ is extremely high, all light elements are destroyed. But as the photon density becomes lower, there is some range of the high energy photon density at which the overproduction of D, T and $^3$He becomes significant. And if the density is sufficiently low, the high energy photon does not affect the BBN at all.

From various observations, the primordial abundances of light elements are estimated (Walker *et al.* 1991) as

$$0.22 < Y_p \equiv \left(\frac{\rho_{^4\text{He}}}{\rho_B}\right)_p < 0.24, \tag{19}$$

$$\left(\frac{n_\text{D}}{n_\text{H}}\right)_p > 1.8 \times 10^{-5}, \tag{20}$$

$$\left(\frac{n_\text{D} + n_{^3\text{He}}}{n_\text{H}}\right)_p < 1.0 \times 10^{-4}, \tag{21}$$

where $\rho_{^4\text{He}}$ and $\rho_B$ are the mass densities of $^4$He and baryon. The abundances of light elements modified by the high energy photons must satisfy the observational constraints above.[3] In order to make precise predictions for the abundances of light elements, we have modified Kawano's computer code (Kawano 1992) to include the photo-dissociation processes.

As an example of the high energy photon source, we study the generic unstable particle $X$ which decays into a photon and a daughter particle ($X \rightarrow \gamma + X'$) (Lindley 1985, Juszkiewicz, Silk and Stebbins 1985, Kawasaki, Terasawa and Sato 1986, Ellis, Nanopoulos and Sarker 1985, Ellis *et al.* 1992), and calculate the abundances of light elements with taking the mass $m_X$, lifetime $\tau_X$ and the abundance $Y_{X0} \equiv n_{X0}/(\zeta(3)T^3/\pi^2)$ of the unstable particle as free parameters ($n_{X0}$ is the number density of the heavy particle before decay). In the BBN calculation, the baryon-to-photon ratio $\eta_B$ is also a free parameter. However as shown in the previous paper (Kawasaki and Moroi 1994a), the baryon-to-photon ratio $\eta_B$ is not an important parameter because the allowed value for $\eta_B$ is almost the same as that in the standard case. Therefore we fix $\eta_B = 3 \times 10^{-10}$ in this paper. Assuming that the unstable particles are almost at rest when it decays and that the emitted high energy photons have

---

[3]In our analysis, we have not considered the constraints from the abundance of $^7$Li, since the cross section data for $^7$Be is not available and hence we cannot predict the abundance of $^7$Li a part of which come from $^7$Be.

unique energy $\epsilon_0 = m_X/2$, $(\partial f_\gamma/\partial t)|_S$ can be written as

$$\left.\frac{\partial f_\gamma(\epsilon_\gamma)}{\partial t}\right|_S = \delta(\epsilon_\gamma - \epsilon_0)\frac{n_X}{\tau_X}. \quad (22)$$

Then, comparing the calculated abundance of the light element with (19)–(21), we can obtain the constraints on the mass, lifetime and abundance of the unstable particle. (For details, see Kawasaki and Moroi 1994a.)

In Figs4 – 6 the constraints are shown for the case $\epsilon_0$ being (a) 10MeV, (b)10GeV and (c) 10TeV in the $\epsilon_0 Y_{X0} - \tau_X$ plane. Since the spectrum of the cascade photons depends only on the energy injection for large $\epsilon_0$, the constraint for $\epsilon_0 = 10$GeV is the almost the same as that for $\epsilon_0 = 10$TeV. As is mentioned before, if the lifetime is short ($\tau_X \lesssim 10^6$ sec), the photon-photon process is effective below the threshold of $^4$He photo-dissociation and only D and $^3$He are destroyed. On the other hand, for long lifetime, the photon-photon process is not so effective that the cascade photons destroy the $^4$He and result in the overproduction of D and $^3$He. When the energy of the injected photon is less than about 10MeV, only D can be destroyed (Fig.4). In any case, the obtained constraints are very stringent, in particular, when the cascade occurs at $t \sim 10^4$sec or later.

## V. Conclusions

We study the electromagnetic cascade in the early Universe. The spectrum of the cascade photons is obtained by finding steady-state-solution to the Boltzmann equations which describe the time evolution of the cascade photons and electrons. As shown in Table 1, the resultant spectra both in low energy region and in the photon-photon scattering region fit to the power low spectra (15) and (16), and the spectrum index in the low energy is $-(1.2 - 1.5)$, which indicates that the cascade spectrum is slightly softer than the spectrum with neglecting Compton scattering off the background electrons. This softening of the spectrum may be due to the dependence of Compton scattering on the photon energy. For high $\epsilon_\gamma$, the cross section of Compton scattering is proportional to $\epsilon_\gamma^{-1}$. This means that the mean free time of the high energy photons becomes longer as their energy increases. Since the amplitude of the spectrum is roughly proportional to the mean free time, Compton scattering relatively increases the amplitude in the high energy region and softens the spectrum.

We also apply our cascade spectrum to BBN calculation with unstable heavy particle and obtain the stringent constraints on the lifetime and the abundance of the heavy particle;

$$\epsilon_0 Y_{X0} \lesssim 10^{-6} - 10^{-9}\text{GeV} \quad \text{for} \quad 10^4 \text{ sec} \lesssim \tau_X \lesssim 10^6 \text{ sec}, \quad (23)$$

$$\epsilon_0 Y_{X0} \lesssim 10^{-13}\text{GeV} \quad \text{for} \quad 10^6 \text{ sec} \lesssim \tau_X, \quad \epsilon_0 \gtrsim 20\text{MeV}, \quad (24)$$

$$\epsilon_0 Y_{X0} \lesssim 10^{-9}\text{GeV} \quad \text{for} \quad 10^6 \text{ sec} \lesssim \tau_X, \quad \epsilon_0 \lesssim 20\text{MeV}. \quad (25)$$

Here we places the constraints on unstable heavy particles. However, if we read the lifetime as the epoch for the injection of the high energy photons, the above constraints apply to generic photon sources.

Finally we briefly comment on the recent related work. Protheroe, Stanev and Berezinsky (1994) calculated the cascade spectrum including Compton scattering and other processes by using complicated Monte Carlo simulations. Since they do not present any spectra, we cannot directly compare our spectra with theirs. However they also obtain constraints on heavy unstable particles, which are consistent with our results.

# Appendix : Boltzmann Equation

To make this paper selfcontained, we show the details of the Boltzmann equations which govern the electromagnetic cascade processes in this appendix.

### (I) DOUBLE PHOTON PAIR CREATION [ $\gamma + \gamma \to e^+ + e^-$ ]

For high energy photon whose energy is larger than $\sim m_e^2/22T$, double photon pair creation is the most dominant process.

The total cross section $\sigma_{\rm DP}$ for the double photon pair creation process is given by

$$\sigma_{\rm DP}(\beta) = \frac{1}{2}\pi r_e^2 \left(1-\beta^2\right)\left\{\left(3-\beta^4\right)\ln\frac{1+\beta}{1-\beta} - 2\beta\left(2-\beta^2\right)\right\}, \tag{26}$$

where $r_e = \alpha/m_e$ is the classical radius of electron and $\beta$ is the electron (or positron) velocity in the center of mass frame. Using this formula, one can write down $(\partial f_\gamma/\partial t)|_{\rm DP}$ as

$$\left.\frac{\partial f_\gamma(\epsilon_\gamma)}{\partial t}\right|_{\rm DP} = -\frac{1}{8}\frac{1}{\epsilon_\gamma^2}f_\gamma(\epsilon_\gamma)\int_{m_e/\epsilon_\gamma}^{\infty}d\bar{\epsilon}_\gamma\frac{1}{\bar{\epsilon}_\gamma^2}\bar{f}_\gamma(\bar{\epsilon}_\gamma)\int_{4m_e^2}^{4\epsilon_\gamma\bar{\epsilon}_\gamma}ds\ s\sigma_{DP}(\beta)\bigg|_{\beta=\sqrt{1-(4m_e^2/s)}}. \tag{27}$$

The spectrum of final state electron and positron is obtained in Agaronyan, Atyan and Najapetyan (1983), and $(\partial f_e/\partial t)|_{\rm DP}$ is given by

$$\left.\frac{\partial f_e(\epsilon_e)}{\partial t}\right|_{\rm DP} = \frac{1}{4}\pi r_e^2 m_e^4 \int_{\epsilon_e}^{\infty}d\epsilon_\gamma\frac{f_\gamma(\epsilon_\gamma)}{\epsilon_\gamma^3}\int_0^{\infty}d\bar{\epsilon}_\gamma\frac{\bar{f}_\gamma(\bar{\epsilon}_\gamma)}{\bar{\epsilon}_\gamma^2}G(\epsilon_e,\epsilon_\gamma,\bar{\epsilon}_\gamma), \tag{28}$$

where $\bar{f}_\gamma$ represents the distribution function of the background photon at temperature $T$,

$$\bar{f}_\gamma(\bar{\epsilon}_\gamma) = \frac{\bar{\epsilon}_\gamma^2}{\pi^2}\times\frac{1}{\exp(\bar{\epsilon}_\gamma/T)-1}, \tag{29}$$

and function $G(\epsilon_e,\epsilon_\gamma,\bar{\epsilon}_\gamma)$ is given by

$$G(\epsilon_e,\epsilon_\gamma,\bar{\epsilon}_\gamma) = \frac{4\left(\epsilon_e+\epsilon_e'\right)^2}{\epsilon_e\epsilon_e'}\ln\frac{4\bar{\epsilon}_\gamma\epsilon_e\epsilon_e'}{m_e^2\left(\epsilon_e+\epsilon_e'\right)} - \left\{1-\frac{m_e^2}{\bar{\epsilon}_\gamma\left(\epsilon_e+\epsilon_e'\right)}\right\}\frac{\left(\epsilon_e+\epsilon_e'\right)^4}{\epsilon_e^2\epsilon_e'^2}$$

$$+\frac{2\left\{2\bar{\epsilon}_\gamma\left(\epsilon_e+\epsilon'_e\right)-m_e^2\right\}\left(\epsilon_e+\epsilon'_e\right)^2}{m_e^2\epsilon_e\epsilon'_e}-8\frac{\bar{\epsilon}_\gamma\epsilon_\gamma}{m_e^2}, \quad (30)$$

with $\epsilon'_e = \epsilon_\gamma + \bar{\epsilon}_\gamma - \epsilon_e$.

## (II) PHOTON-PHOTON SCATTERING [ $\gamma + \gamma \to \gamma + \gamma$ ]

If the photon energy is below the effective threshold of the double photon pair creation, photon-photon scattering process becomes significant. This process is analyzed in Svensson and Zdziarski (1990) and for $\epsilon'_\gamma \lesssim O(m_e^2/T)$, $(\partial f_\gamma/\partial t)|_{\rm PP}$ is given by

$$\begin{aligned}\left.\frac{\partial f_\gamma(\epsilon'_\gamma)}{\partial t}\right|_{\rm PP} &= \frac{35584}{10125\pi}\alpha^2 r_e^2 m_e^{-6}\int_{\epsilon'_\gamma}^\infty d\epsilon_\gamma f_\gamma(\epsilon_\gamma)\epsilon_\gamma^2\left\{1-\frac{\epsilon'_\gamma}{\epsilon_\gamma}+\left(\frac{\epsilon'_\gamma}{\epsilon_\gamma}\right)^2\right\}^2\int_0^\infty d\bar{\epsilon}_\gamma\bar{\epsilon}_\gamma^3\bar{f}_\gamma(\bar{\epsilon}_\gamma)\\
&\quad -\frac{1946}{50625\pi}f_\gamma(\epsilon'_\gamma)\alpha^2 r_e^2 m_e^{-6}{\epsilon'_\gamma}^3\int_0^\infty d\bar{\epsilon}_\gamma\bar{\epsilon}_\gamma^3\bar{f}_\gamma(\bar{\epsilon}_\gamma).\end{aligned} \quad (31)$$

For a larger value of $\epsilon'_\gamma$, we cannot use this formula. But for high energy photons, photon-photon scattering is not significant because double photon pair creation determines the shape of the photon spectrum. Therefore, instead of using the exact formula, we have taken $m_e^2/T$ as a cutoff scale of $(\partial f_\gamma/\partial t)|_{\rm PP}$, i.e. for $\epsilon'_\gamma \leq m_e^2/T$ we have used eq.(31) and for $\epsilon'_\gamma > m_e^2/T$ we have taken

$$\left.\frac{\partial f_\gamma(\epsilon'_\gamma > m_e^2/T)}{\partial t}\right|_{\rm PP} = 0. \quad (32)$$

Notice that we have checked the cutoff dependence of spectra is negligible.

## (III) PAIR CREATION IN NUCLEI [ $\gamma + N \to e^+ + e^- + N$ ]

Scattering off the electric field around nucleon, high energy photon can produce electron positron pair if the photon energy is larger than $2m_e$. Denoting total cross section of this process $\sigma_{\rm PC}$, $(\partial f_\gamma/\partial t)|_{\rm NP}$ is given by

$$\left.\frac{\partial f_\gamma(\epsilon_\gamma)}{\partial t}\right|_{\rm NP} = -n_N\sigma_{\rm PC}(\epsilon_\gamma)f_\gamma(\epsilon_\gamma), \quad (33)$$

where $n_N$ is the nucleon number density. For $\sigma_{\rm PC}$, we have used the approximate formula derived by Maximon (1968). For $\epsilon_\gamma$ near the threshold ($\epsilon_\gamma < 4m_e$), the approximate formula is given by

$$\sigma_{\rm PC}(\epsilon_\gamma)|_{k<4} = Z^2\alpha r_e^2\frac{2\pi}{3}\left(\frac{k-2}{k}\right)^3\left(1+\frac{1}{2}\rho+\frac{23}{40}\rho^2+\frac{11}{60}\rho^3+\frac{29}{960}\rho^4+O(\rho^5)\right), \quad (34)$$

where

$$k \equiv \frac{\epsilon_\gamma}{m_e}, \quad \rho \equiv \frac{2k-4}{k+2+2\sqrt{2k}}, \qquad (35)$$

and $Z$ is the charge of nuclei. For large $\epsilon_\gamma$ ($\epsilon_\gamma \geq 4m_e$), $\sigma_{\rm PC}$ are expanded in the parameter $k^{-1}$, which is given by

$$\begin{aligned}
\sigma_{\rm PC}(\epsilon_\gamma)|_{k\geq 4} &= Z^2 \alpha r_e^2 \bigg[ \frac{28}{9} \ln 2k - \frac{218}{27} \\
&\quad + \left(\frac{2}{k}\right)^2 \left\{ \frac{2}{3} (\ln 2k)^3 - (\ln 2k)^2 + \left(6 - \frac{\pi^2}{3}\right) \ln 2k + 2\zeta(3) + \frac{\pi^2}{6} - \frac{7}{2} \right\} \\
&\quad - \left(\frac{2}{k}\right)^4 \left\{ \frac{3}{16} \ln 2k + \frac{1}{8} \right\} \\
&\quad - \left(\frac{2}{k}\right)^6 \left\{ \frac{29}{2304} \ln 2k - \frac{77}{13824} \right\} + O(k^{-8}) \bigg].
\end{aligned} \qquad (36)$$

Differential cross section for this process $d\sigma_{\rm PC}/dE_+$ is given in Berestetskiĭ, Lifshitz and Pitaevskiĭ (1971);

$$\begin{aligned}
\frac{d\sigma_{\rm PC}}{dE_+} &= Z^2 \alpha r_e^2 \left(\frac{p_+ p_-}{\epsilon_\gamma^3}\right) \bigg[ -\frac{4}{3} - 2E_+ E_- \frac{p_+^2 + p_-^2}{p_+^2 p_-^2} \\
&\quad + m_e^2 \left\{ l_- \frac{E_+}{p_-^3} + l_+ \frac{E_-}{p_+^3} - l_+ l_- \frac{1}{p_+ p_-} \right\} \\
&\quad + L \bigg\{ -\frac{8 E_+ E_-}{3 p_+ p_-} + \frac{\epsilon_\gamma^2}{p_+^3 p_-^3} \left(E_+^2 E_-^2 + p_+^2 p_-^2 - m_e^2 E_+ E_-\right) \\
&\quad - \frac{m_e^2 \epsilon_\gamma}{2 p_+ p_-} \left( l_+ \frac{E_+ E_- - p_+^2}{p_+^3} + l_- \frac{E_+ E_- - p_-^2}{p_-^3} \right) \bigg\} \bigg],
\end{aligned} \qquad (37)$$

where

$$p_\pm \equiv \sqrt{E_\pm^2 - m_e^2}, \qquad (38)$$

$$L \equiv \ln \frac{E_+ E_- + p_+ p_- + m_e^2}{E_+ E_- - p_+ p_- + m_e^2}, \qquad (39)$$

$$l_\pm \equiv \ln \frac{E_\pm + p_\pm}{E_\pm - p_\pm}, \qquad (40)$$

with $E_-$ ($E_+$) being the energy of electron (positron) in final state. By using this formula, $(\partial f_e/\partial t)|_{\rm NP}$ is given by

$$\left.\frac{\partial f_e(\epsilon_e)}{\partial t}\right|_{\rm NP} = n_N \int_{\epsilon_e+m_e}^\infty d\epsilon_\gamma \frac{d\sigma_{\rm PC}}{d\epsilon_e} f_\gamma(\epsilon_\gamma). \qquad (41)$$

## (IV) COMPTON SCATTERING [ $\gamma + e^- \to \gamma + e^-$ ]

Compton scattering is one of the processes by which high energy photons lose their energy. Since the photo-dissociation of light elements occurs when the temperature drops below $\sim 0.1$MeV, we can consider the thermal electrons to be almost at rest. Using the total and differential cross sections at the electron rest frame $\sigma_{\rm CS}$ and $d\sigma_{\rm CS}/d\epsilon_e$, one can derive

$$\left.\frac{\partial f_\gamma(\epsilon'_\gamma)}{\partial t}\right|_{\rm CS} = \bar{n}_e \int_{\epsilon'_\gamma}^\infty d\epsilon_\gamma f_\gamma(\epsilon_\gamma) \frac{d\sigma_{\rm CS}(\epsilon'_\gamma, \epsilon_\gamma)}{d\epsilon'_\gamma} - \bar{n}_e \sigma_{\rm CS} f_\gamma(\epsilon'_\gamma), \tag{42}$$

$$\left.\frac{\partial f_e(\epsilon'_e)}{\partial t}\right|_{\rm CS} = \bar{n}_e \int_{\epsilon'_e}^\infty d\epsilon_\gamma f_\gamma(\epsilon_\gamma) \frac{d\sigma_{\rm CS}(\epsilon_\gamma + m_e - \epsilon'_e, \epsilon_\gamma)}{d\epsilon'_\gamma}, \tag{43}$$

where $\bar{n}_e$ is the number density of background electron. Explicit form of $\sigma_{\rm CS}$ and $d\sigma_{\rm CS}/d\epsilon_e$ are given as

$$\sigma_{\rm CS} = 2\pi r_e^2 \frac{1}{x}\left\{\left(1 - \frac{4}{x} - \frac{8}{x^2}\right)\ln(1+x) + \frac{1}{2} + \frac{8}{x} - \frac{1}{2(1+x)^2}\right\}, \tag{44}$$

$$\frac{d\sigma_{\rm CS}(\epsilon'_\gamma, \epsilon_\gamma)}{d\epsilon'_\gamma} = \pi r_e^2 \frac{m_e}{\epsilon_\gamma^2}\left\{\frac{\epsilon_\gamma}{\epsilon'_\gamma} + \frac{\epsilon'_\gamma}{\epsilon_\gamma} + \left(\frac{m_e}{\epsilon'_\gamma} - \frac{m_e}{\epsilon_\gamma}\right)^2 - 2m_e\left(\frac{1}{\epsilon'_\gamma} - \frac{1}{\epsilon_\gamma}\right)\right\}, \tag{45}$$

with

$$x \equiv \frac{s - m_e^2}{m_e^2} = \frac{2\epsilon_\gamma}{m_e}. \tag{46}$$

## (V) INVERSE COMPTON SCATTERING [ $e^\pm + \gamma \to e^\pm + \gamma$ ]

Contribution from the inverse Compton process is given by Jones (1968), and $(\partial f/\partial t)|_{\rm IC}$ is given by

$$\left.\frac{\partial f_\gamma(\epsilon_\gamma)}{\partial t}\right|_{\rm IC} = 2\pi r_e^2 m_e^2 \int_{\epsilon_\gamma + m_e}^\infty d\epsilon_e \frac{2 f_e(\epsilon_e)}{\epsilon_e^2} \int_0^\infty d\bar{\epsilon}_\gamma \frac{\bar{f}_\gamma(\bar{\epsilon}_\gamma)}{\bar{\epsilon}_\gamma} F(\epsilon_\gamma, \epsilon_e, \bar{\epsilon}_\gamma), \tag{47}$$

$$\left.\frac{\partial f_e(\epsilon'_e)}{\partial t}\right|_{\rm IC} = 2\pi r_e^2 m_e^2 \int_{\epsilon'_e}^\infty d\epsilon_e \frac{f_e(\epsilon_e)}{\epsilon_e^2} \int_0^\infty d\bar{\epsilon}_\gamma \frac{\bar{f}_\gamma(\bar{\epsilon}_\gamma)}{\bar{\epsilon}_\gamma} F(\epsilon_e + \bar{\epsilon}_\gamma - \epsilon'_e, \epsilon_e, \bar{\epsilon}_\gamma)$$
$$-2\pi r_e^2 m_e^2 \frac{f_e(\epsilon'_e)}{{\epsilon'_e}^2} \int_{\epsilon'_e}^\infty d\epsilon_\gamma \int_0^\infty d\bar{\epsilon}_\gamma \frac{\bar{f}_\gamma(\bar{\epsilon}_\gamma)}{\bar{\epsilon}_\gamma} F(\epsilon_\gamma, \epsilon'_e, \bar{\epsilon}_\gamma), \tag{48}$$

where function $F(\epsilon_\gamma, \epsilon_e, \bar{\epsilon}_\gamma)$ is given by

$$F(\epsilon_\gamma, \epsilon_e, \bar{\epsilon}_\gamma)|_{0 \geq q \geq 1} = 2q \ln q + (1 + 2q)(1 - q) + \frac{(\Gamma_\epsilon q)^2}{2(1 - \Gamma_\epsilon q)}(1 - q), \tag{49}$$

$$F(\epsilon_\gamma, \epsilon_e, \bar{\epsilon}_\gamma)|_{\rm otherwise} = 0, \tag{50}$$

with

$$\Gamma_\epsilon = \frac{4\bar{\epsilon}_\gamma \epsilon_e}{m_e^2}, \quad q = \frac{\epsilon_\gamma}{\Gamma_\epsilon(\epsilon_e - \epsilon_\gamma)}.$$

$\epsilon_{\gamma 0} = 10$ TeV

| Temperature | $P_{\text{low}}$ | $N_{\text{low}}$ GeV$^2$ | $P_{\text{pp}}$ | $N_{\text{pp}}$ GeV$^2$ |
|---:|---|---|---|---|
| 1 eV | $-1.57$ | $1.6 \times 10^8$ | $-5.10$ | $6.9 \times 10^{-18}$ |
| 10 eV | $-1.34$ | $5.4 \times 10^8$ | $-5.20$ | $6.0 \times 10^{-18}$ |
| 100 eV | $-1.22$ | $1.7 \times 10^9$ | $-4.84$ | $1.1 \times 10^{-17}$ |

$\epsilon_{\gamma 0} = 1$ TeV

| Temperature | $P_{\text{low}}$ | $N_{\text{low}}$ GeV$^2$ | $P_{\text{pp}}$ | $N_{\text{pp}}$ GeV$^2$ |
|---:|---|---|---|---|
| 1 eV | $-1.56$ | $1.4 \times 10^8$ | $-5.07$ | $6.2 \times 10^{-18}$ |
| 10 eV | $-1.34$ | $4.9 \times 10^8$ | $-5.17$ | $5.5 \times 10^{-18}$ |
| 100 eV | $-1.22$ | $1.4 \times 10^9$ | $-4.79$ | $1.0 \times 10^{-17}$ |

$\epsilon_{\gamma 0} = 100$ GeV

| Temperature | $P_{\text{low}}$ | $N_{\text{low}}$ GeV$^2$ | $P_{\text{pp}}$ | $N_{\text{pp}}$ GeV$^2$ |
|---:|---|---|---|---|
| 1 eV | $-1.56$ | $1.4 \times 10^8$ | $-5.01$ | $5.7 \times 10^{-18}$ |
| 10 eV | $-1.33$ | $4.7 \times 10^8$ | $-5.15$ | $5.3 \times 10^{-18}$ |
| 100 eV | $-1.22$ | $1.3 \times 10^9$ | $-4.74$ | $1.1 \times 10^{-17}$ |

$\epsilon_{\gamma 0} = 10$ GeV

| Temperature | $P_{\text{low}}$ | $N_{\text{low}}$ GeV$^2$ | $P_{\text{pp}}$ | $N_{\text{pp}}$ GeV$^2$ |
|---:|---|---|---|---|
| 1 eV | — | — | — | — |
| 10 eV | $-1.33$ | $4.5 \times 10^8$ | $-5.12$ | $5.5 \times 10^{-18}$ |
| 100 eV | $-1.22$ | $1.3 \times 10^9$ | $-4.77$ | $9.6 \times 10^{-18}$ |

Table 1: $P_{\text{low}}$, $N_{\text{low}}$, $P_{\text{pp}}$ and $N_{\text{pp}}$ for the cases of $T = 1$ eV, 10 eV, 100 eV, and $\epsilon_0 = 10$TeV, 1TeV, 100GeV, 10GeV.

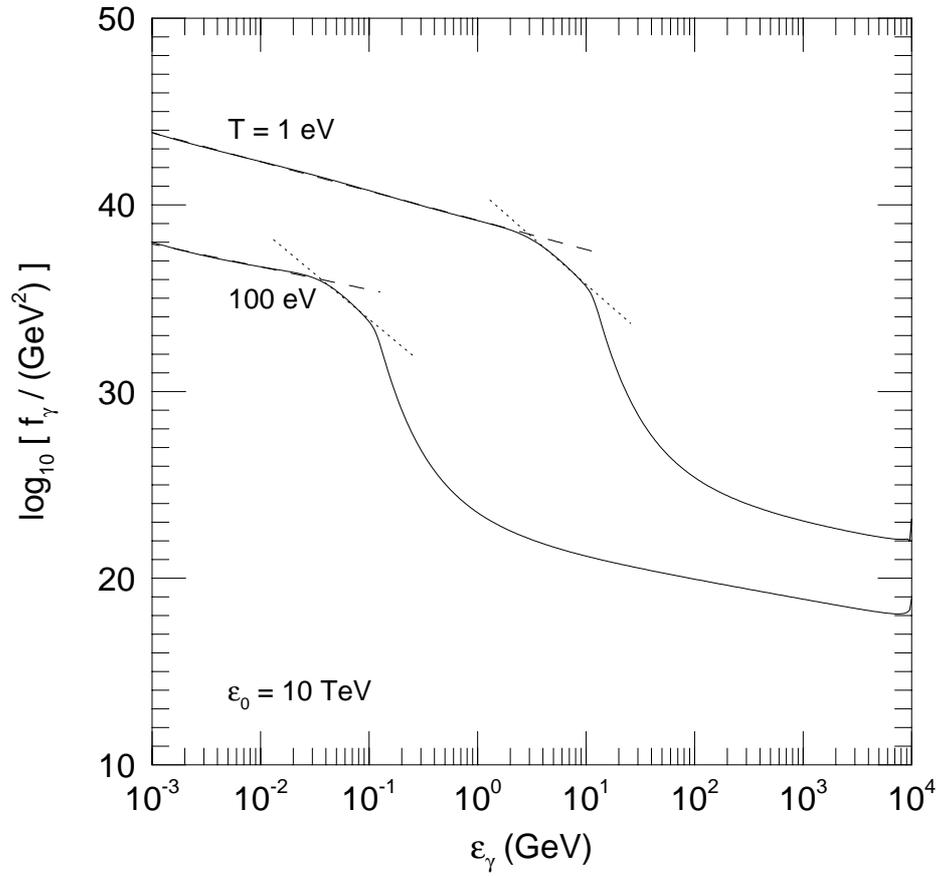

Figure 1: Spectra of the cascade photons for $\epsilon_0 = 10\,\mathrm{TeV}$. In the figure the spectra is shown for $T = 1$ eV and 100 eV. Dashed and dotted lines show the fit by power-low spectrum.

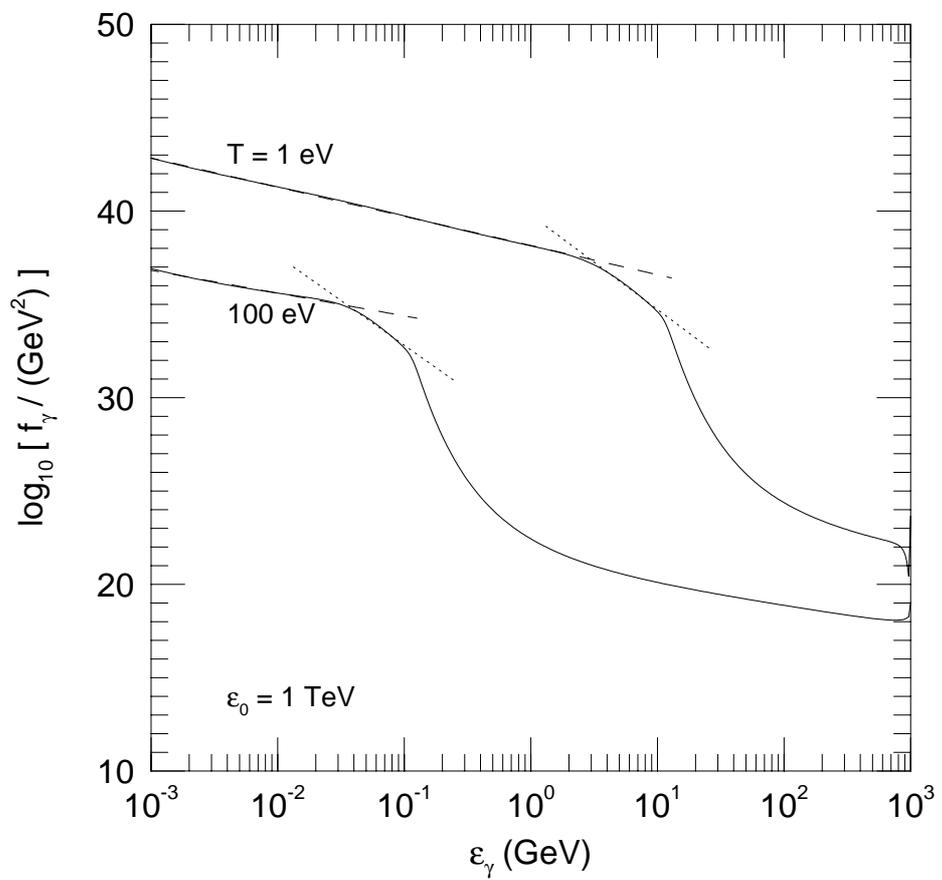

Figure 2: Same as Fig.1 except for $\epsilon_0 = 1\text{TeV}$.

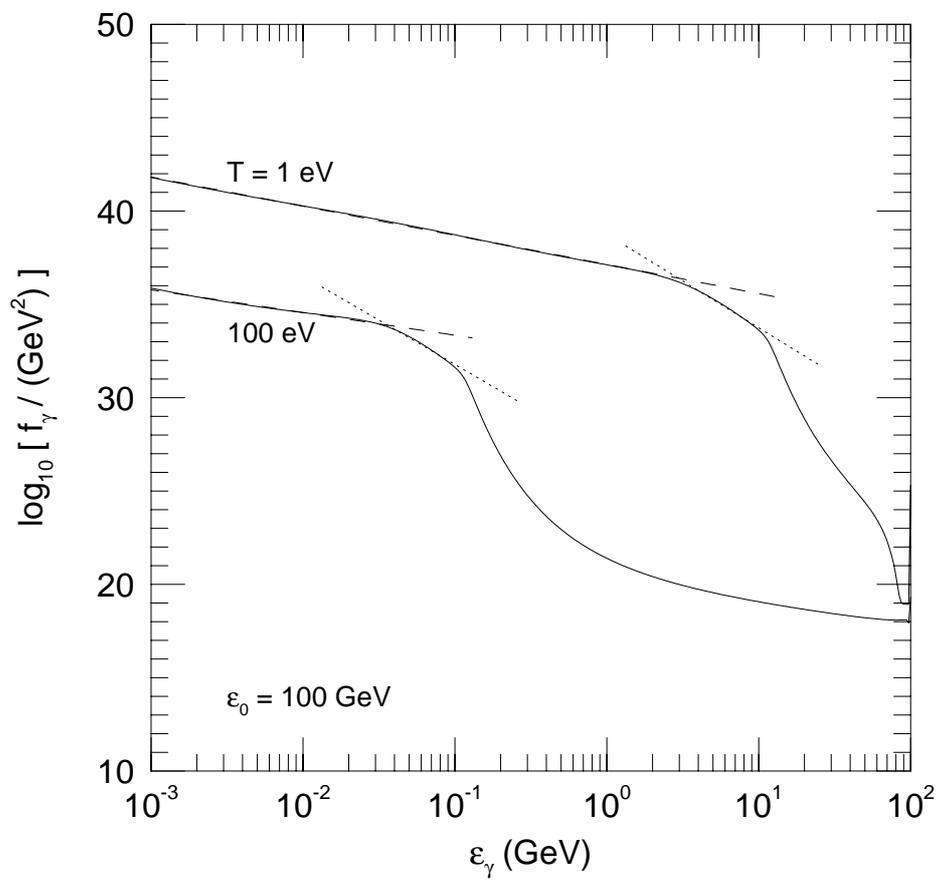

Figure 3: Same as Fig.1 except for $\epsilon_0 = 100$ GeV.

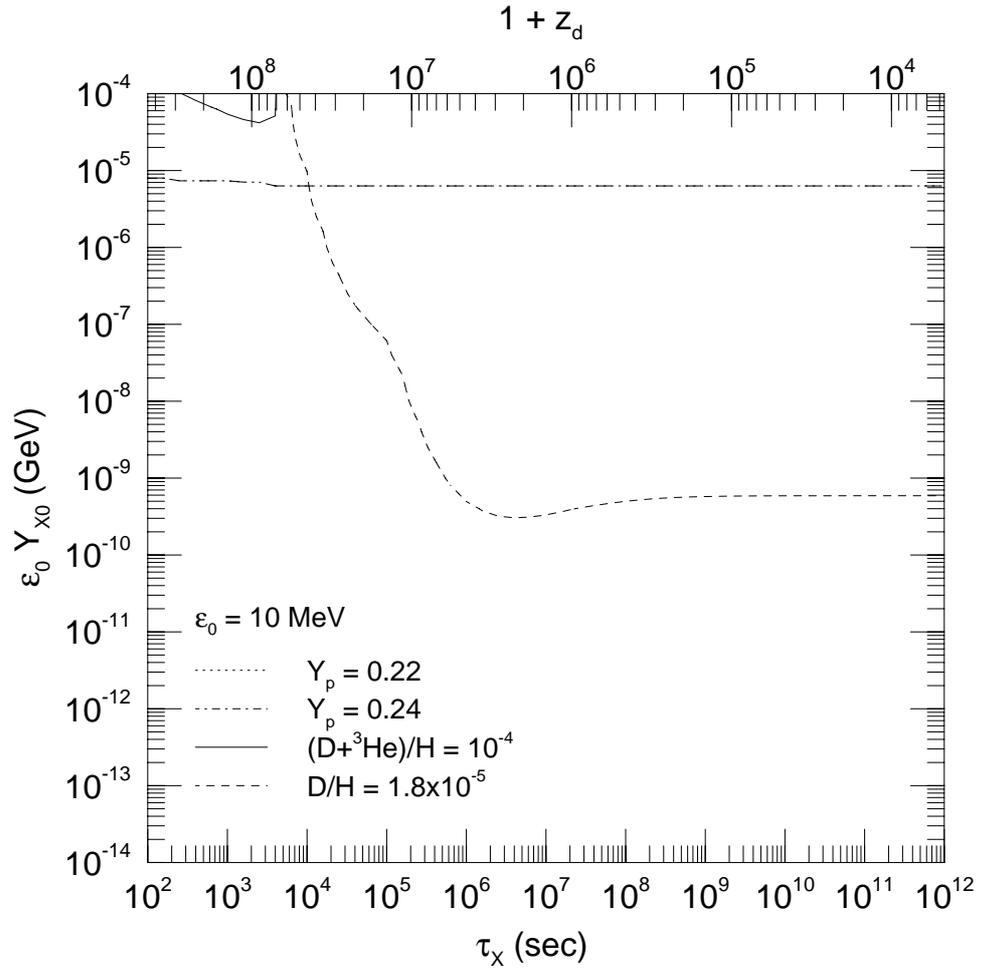

Figure 4: Constraints on the lifetime and the abundance of the unstable particle from BBN for $\epsilon_0 = 10$MeV. Top horizontal axis shows the redshift $z_d$ corresponding to $\tau_X$.

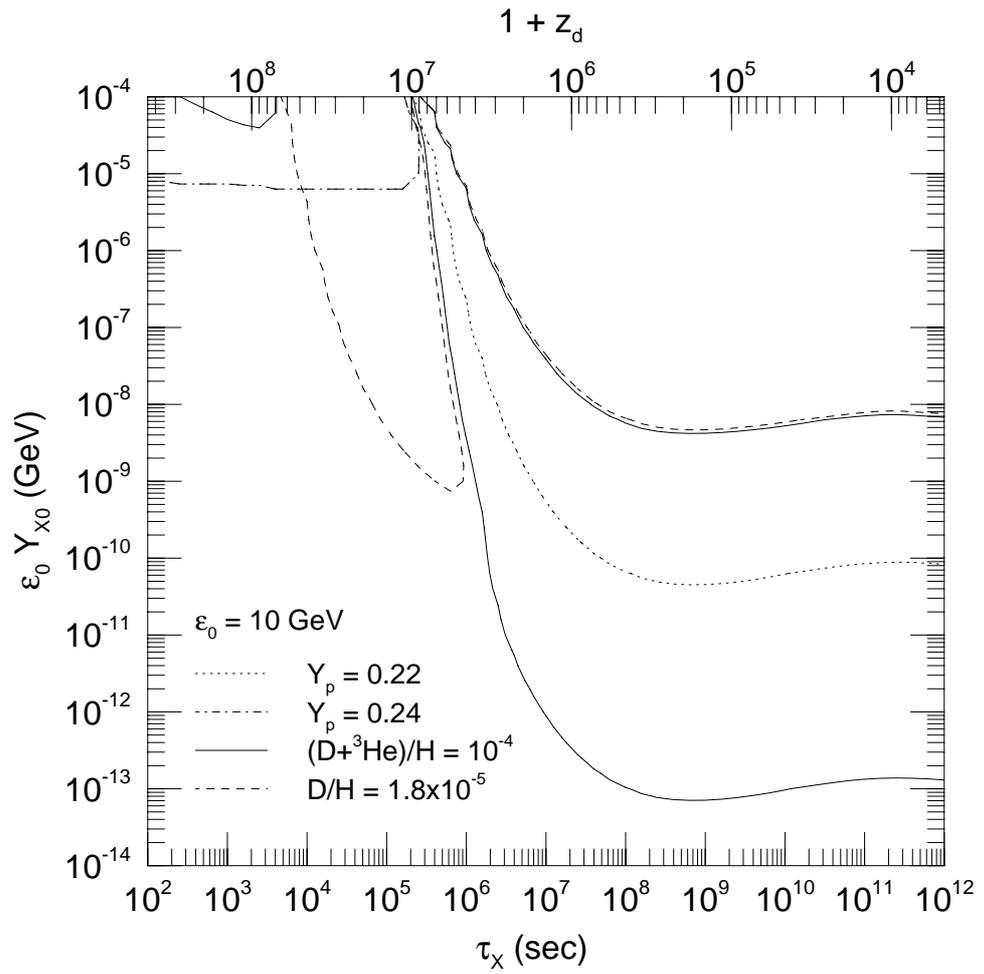

Figure 5: Same as Fig.4 except for $\epsilon_0 = 10\text{GeV}$.

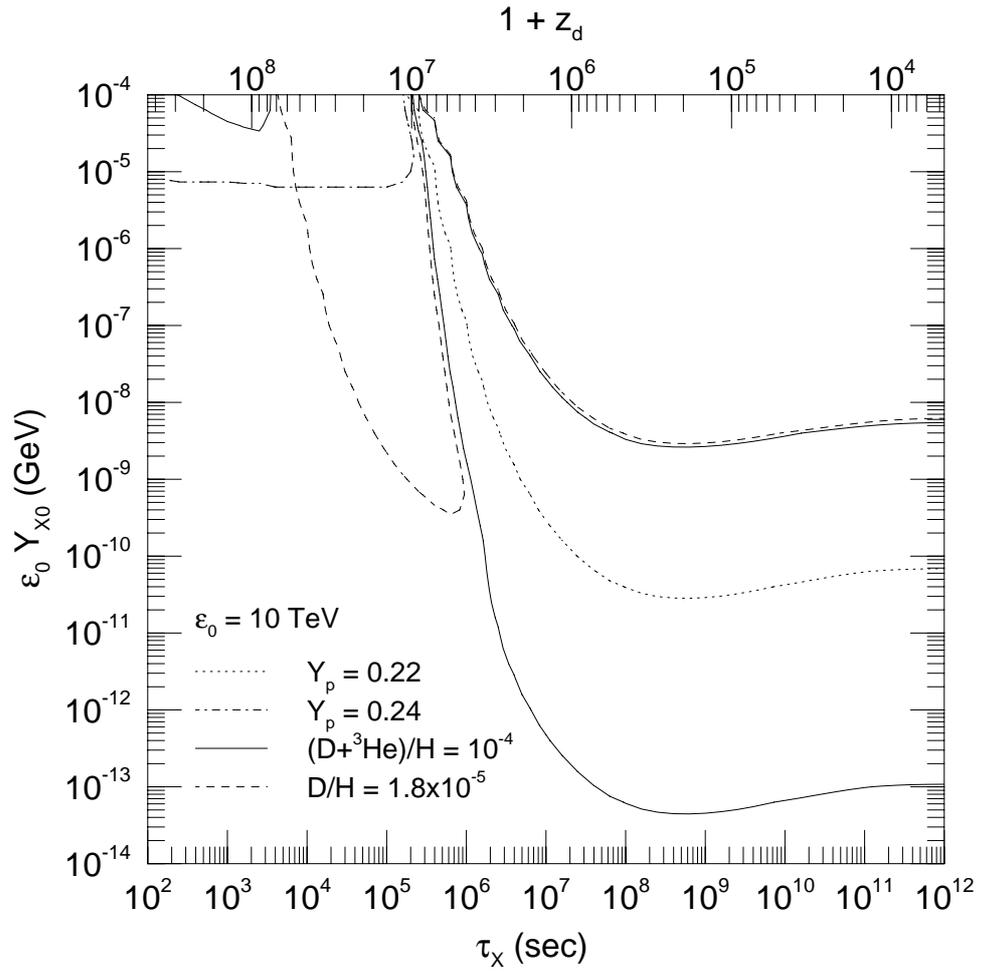

Figure 6: Same as Fig.4 except for $\epsilon_0 = 10\text{TeV}$.

20